\documentclass[pra,aps,nopacs,onecolumn,twoside,superscriptaddress]{revtex4}
\usepackage{amsmath,amsfonts,amssymb,caption,color,epsfig,graphics,graphicx,hyperref,latexsym,mathrsfs,revsymb,theorem,url,verbatim,epstopdf,mathtools,enumerate,tikz}
\usetikzlibrary{matrix}
\usetikzlibrary{decorations.pathreplacing}
\hypersetup{colorlinks,linkcolor={blue},citecolor={blue},urlcolor={red}}

\newtheorem{definition}{Definition}
\newtheorem{proposition}[definition]{Proposition}
\newtheorem{lemma}[definition]{Lemma}

\newtheorem{theorem}[definition]{Theorem}
\newtheorem{corollary}[definition]{Corollary}
\newtheorem{conjecture}[definition]{Conjecture}

\newtheorem{remark}[definition]{Remark}
\newtheorem{example}[definition]{Example}
\newtheorem{question}[definition]{Question}
\newtheorem{memo}[definition]{Memo}

\def\squareforqed{\hbox{\rlap{$\sqcap$}$\sqcup$}}
\def\qed{\ifmmode\squareforqed\else{\unskip\nobreak\hfil
\penalty50\hskip1em\null\nobreak\hfil\squareforqed
\parfillskip=0pt\finalhyphendemerits=0\endgraf}\fi}
\def\endenv{\ifmmode\;\else{\unskip\nobreak\hfil
\penalty50\hskip1em\null\nobreak\hfil\;
\parfillskip=0pt\finalhyphendemerits=0\endgraf}\fi}
\newenvironment{proof}{\noindent \textbf{{Proof.~} }}{\qed}
\def\Dbar{\leavevmode\lower.6ex\hbox to 0pt
{\hskip-.23ex\accent"16\hss}D}
\makeatletter
\def\url@leostyle{%
  \@ifundefined{selectfont}{\def\UrlFont{\sf}}{\def\UrlFont{\small\ttfamily}}}
\makeatother
\def\bcj{\begin{conjecture}}
\def\ecj{\end{conjecture}}
\def\bcr{\begin{corollary}}
\def\ecr{\end{corollary}}
\def\bd{\begin{definition}}
\def\ed{\end{definition}}
\def\bea{\begin{eqnarray}}
\def\eea{\end{eqnarray}}
\def\beq{\begin{equation}}
\def\eeq{\end{equation}}
\def\bal{\begin{aligned}}
\def\eal{\end{aligned}}
\def\bem{\begin{enumerate}}
\def\eem{\end{enumerate}}
\def\bex{\begin{example}}
\def\eex{\end{example}}
\def\bim{\begin{itemize}}
\def\eim{\end{itemize}}
\def\bl{\begin{lemma}}
\def\el{\end{lemma}}
\def\bma{\begin{bmatrix}}
\def\ema{\end{bmatrix}}
\def\bpf{\begin{proof}}
\def\epf{\end{proof}}
\def\bpp{\begin{proposition}}
\def\epp{\end{proposition}}
\def\bqu{\begin{question}}
\def\equ{\end{question}}
\def\br{\begin{remark}}
\def\er{\end{remark}}
\def\bt{\begin{theorem}}
\def\et{\end{theorem}}
\def\bmm{\begin{memo}}
\def\emm{\end{memo}}

\def\btb{\begin{tabular}}
\def\etb{\end{tabular}}

\newcommand{\nc}{\newcommand}

\def\a{\alpha}
\def\b{\beta}

\nc{\bbA}{\mathbb{A}} \nc{\bbB}{\mathbb{B}} \nc{\bbC}{\mathbb{C}}
 \nc{\bbD}{\mathbb{D}} \nc{\bbE}{\mathbb{E}} \nc{\bbF}{\mathbb{F}}
 \nc{\bbG}{\mathbb{G}} \nc{\bbH}{\mathbb{H}} \nc{\bbI}{\mathbb{I}}
 \nc{\bbJ}{\mathbb{J}} \nc{\bbK}{\mathbb{K}} \nc{\bbL}{\mathbb{L}}
 \nc{\bbM}{\mathbb{M}} \nc{\bbN}{\mathbb{N}} \nc{\bbO}{\mathbb{O}}
 \nc{\bbP}{\mathbb{P}} \nc{\bbQ}{\mathbb{Q}} \nc{\bbR}{\mathbb{R}}
 \nc{\bbS}{\mathbb{S}} \nc{\bbT}{\mathbb{T}} \nc{\bbU}{\mathbb{U}}
 \nc{\bbV}{\mathbb{V}} \nc{\bbW}{\mathbb{W}} \nc{\bbX}{\mathbb{X}}
 \nc{\bbZ}{\mathbb{Z}}

 \nc{\bA}{{\bf A}} \nc{\bB}{{\bf B}} \nc{\bC}{{\bf C}}
 \nc{\bD}{{\bf D}} \nc{\bE}{{\bf E}} \nc{\bF}{{\bf F}}
 \nc{\bG}{{\bf G}} \nc{\bH}{{\bf H}} \nc{\bI}{{\bf I}}
 \nc{\bJ}{{\bf J}} \nc{\bK}{{\bf K}} \nc{\bL}{{\bf L}}
 \nc{\bM}{{\bf M}} \nc{\bN}{{\bf N}} \nc{\bO}{{\bf O}}
 \nc{\bP}{{\bf P}} \nc{\bQ}{{\bf Q}} \nc{\bR}{{\bf R}}
 \nc{\bS}{{\bf S}} \nc{\bT}{{\bf T}} \nc{\bU}{{\bf U}}
 \nc{\bV}{{\bf V}} \nc{\bW}{{\bf W}} \nc{\bX}{{\bf X}}
 \nc{\bZ}{{\bf Z}}

\nc{\cA}{{\cal A}} \nc{\cB}{{\cal B}} \nc{\cC}{{\cal C}}
\nc{\cD}{{\cal D}} \nc{\cE}{{\cal E}} \nc{\cF}{{\cal F}}
\nc{\cG}{{\cal G}} \nc{\cH}{{\cal H}} \nc{\cI}{{\cal I}}
\nc{\cJ}{{\cal J}} \nc{\cK}{{\cal K}} \nc{\cL}{{\cal L}}
\nc{\cM}{{\cal M}} \nc{\cN}{{\cal N}} \nc{\cO}{{\cal O}}
\nc{\cP}{{\cal P}} \nc{\cQ}{{\cal Q}} \nc{\cR}{{\cal R}}
\nc{\cS}{{\cal S}} \nc{\cT}{{\cal T}} \nc{\cU}{{\cal U}}
\nc{\cV}{{\cal V}} \nc{\cW}{{\cal W}} \nc{\cX}{{\cal X}}
\nc{\cZ}{{\cal Z}}

\nc{\hA}{{\hat{A}}} \nc{\hB}{{\hat{B}}} \nc{\hC}{{\hat{C}}}
\nc{\hD}{{\hat{D}}} \nc{\hE}{{\hat{E}}} \nc{\hF}{{\hat{F}}}
\nc{\hG}{{\hat{G}}} \nc{\hH}{{\hat{H}}} \nc{\hI}{{\hat{I}}}
\nc{\hJ}{{\hat{J}}} \nc{\hK}{{\hat{K}}} \nc{\hL}{{\hat{L}}}
\nc{\hM}{{\hat{M}}} \nc{\hN}{{\hat{N}}} \nc{\hO}{{\hat{O}}}
\nc{\hP}{{\hat{P}}} \nc{\hR}{{\hat{R}}} \nc{\hS}{{\hat{S}}}
\nc{\hT}{{\hat{T}}} \nc{\hU}{{\hat{U}}} \nc{\hV}{{\hat{V}}}
\nc{\hW}{{\hat{W}}} \nc{\hX}{{\hat{X}}} \nc{\hZ}{{\hat{Z}}}

\nc{\hn}{{\hat{n}}}

\def\diag{\mathop{\rm diag}}

\def\max{\mathop{\rm max}}

\usepackage{hyperref}

\begin{document}

\Large

\title{Order-six CHMs containing exactly three distinct elements}

\author{Yanzu Huang}\email[]{21377273@buaa.edu.cn}
\affiliation{LMIB(Beihang University), Ministry of education, and School of Mathematical Sciences, Beihang University, Beijing 100191, China}

\author{Mengfan Liang}\email[]{}
\affiliation{LMIB(Beihang University), Ministry of education, and School of Mathematical Sciences, Beihang University, Beijing 100191, China}

\author{Lin Chen}\email[]{linchen@buaa.edu.cn (corresponding author)}
\affiliation{LMIB(Beihang University), Ministry of education, and School of Mathematical Sciences, Beihang University, Beijing 100191, China}

\begin{abstract}
Complex Hadamard matrices (CHMs) are intimately related to the number of distinct matrix elements. 
We investigate CHMs containing exactly three distinct elements, which is also the least number of distinct elements. In this paper, we show that such CHMs can only be complex equivalent to two kind of matrices, one is $H_2$-reducible and the other is the Tao matrix. Using our result one can further narrow the range of MUB trio
 (a set of four MUBs in $\mathbb{C}^6$ consists of an MUB trio and the identity) since we f{}ind that the two CHMs neither belong to MUB trios.  Our results may
 lead to the more complete  classif{}ication of $6\times 6$ CHMs whose elements in the f{}irst row are all 1.
\end{abstract}

\date{\today}
\maketitle

\section{Intrduction}
\label{sec:int}
Mutually unbiased bases (MUBs) are a signif{}icant concept in quantum physics. In general, MUBs in
Hilbert space $\mathbb{C}^n$ are orthogonal bases such that the inner product of any two vectors from different bases has a modulus of $\dfrac{1}{\sqrt{n}}$. When the number of MUBs reaches $n+1$, they are referred to as complete MUBs. Complete MUBs exist in $\mathbb{C}^n$ when $n$ is a prime power \cite{WOOTTERS1989363}. The problem of f{}inding complete MUBs in $\mathbb{C}^6$ is an unsolved case and a well-known open problem in quantum information. Various approaches have been used to study the MUB problem. For instance, paper \cite{rle11} explored the average distance between four bases in six dimensions, providing strong evidence against the existence of four mutually unbiased bases in $\mathbb{C}^d$. Paper \cite{jmm09} introduced an inf{}inite family of MUB triplets in dimension 6, demonstrating that this family cannot be extended to complete MUBs. Paper \cite{mw12jpa102001} showed that if complete MUBs in dimension 6 exist, they cannot include more than one product basis. Paper \cite{Chen2018Mutually} examined the number of product vectors in a set of four MUBs in dimension 6, showing that each of the remaining three MUBs contains at most two product vectors. Further research on this topic can be found in \cite{Designolle2018Quantifying,mw12jpa135307,mpw16,Boykin05,bw08,bw10,deb10,wpz11,mw12ijqi,Chen2017Product}.\par

The complex Hadamard matrix (CHM) is also an important concept since it is usually related to MUB problem. An $n\times n$ matrix $H$ whose elements all have modulus one is called a CHM if $HH^{\dagger}=nI$. If a set of four MUBs in $\mathbb{C}^6$ exists and contains the identity, then any other matrix $U$ in the set satisf{}ies that $\sqrt{6}U$ is a $6\times 6$ CHM, and we refer to the set as an MUB trio. The f{}inal target for us is to f{}ind if there exists a MUB trio.

The complete classif{}ication of $6\times 6$ CHMs is also a longstanding open problem. Paper \cite{Butson} characterized CHMs whose all elements are roots of unity, such as Fourier matrices. Paper \cite{Taom} introduced the Tao matrix, consisting only of $1,\mathrm{e}^{\frac{2\pi \mathrm{i}}{3}}, \mathrm{e}^{\frac{4\pi \mathrm{i}}{3}}$, which does not belong to any parameterized family of matrices. Paper \cite{BjrckGran} presented a method to apply faster algorithms for the homogeneous case to the inhomogeneous case, discovering a family of $6\times 6$ CHMs not included in the Butson matrices. In 2011, Karlsson introduced a three-parameter family of CHMs in $\mathbb{C}^6$\cite{karlsson11}, termed "the $H_2$-reducible matrices." The most known
$6\times 6$ CHMs, such as the Haagerup matrix \cite{Haagerup1997}, belong to this family, except for the Tao matrix. Paper \cite{Sz12} proposed a four-parameter
$6\times 6$ CHM family, though its analytic form remains unknown. Additional studies on the classif{}ication problem are available in \cite{2006Orthogonal,Tadej2006,Banica2009,Launey2001,DitaP2002,Hiranandani2014}.

In this paper, we investigate the CHM of order 6 containing only three distinct elements. In Theorem \ref{thm:(1,-1,a)}, we claim that the CHM containing only $\{1,-1,a\}$ is complex equivalent to $H^{(1)}$. In Corollary \ref{cr:1-1i-i}, we show that an order-six CHM containing only elements $\{1,-1,i,-i\}$ is complex equivalent to $H^{(1)}$ in \eqref{eq:(1,-1,a)}.  We extend Lemma \ref{lemH2} to f{}inish our classif{}ication of all $H_2$-reducible CHM containing only three elements, and the result is that $H_2$-reducible CHM containing only three elements is complex equivalent to $H^{(1)}$. Next, in Theorem \ref{thm2}, we claim that CHM containing only $\{1,a,\bar{a}\}$ is complex equivalent to $S_6^{(0)}$, the Tao matrix. In Lemma \ref{lem2} we introduce a method to roughly but eff{}iciently classify all the cases. The core idea of that is to preliminarily examine the real part of the inner product of two rows or columns, and that method will be used throughout all subsequent discussions. In Theorem \ref{thm3}, we claim that CHM containing only $\{1,a,-\bar{a}\}$ does not exist. In Lemma \ref{lem3} we extend our method to do a brief classif{}ication, we def{}ine the modif{}ied pending terms, which is somewhat different from several terms in the inner product. Theorem \ref{thmbest} shows the main result in this paper, that is, CHM containing only three distinct elements is complex equivalent to $S_6^{(0)}$ or $H^{(1)}$. We prove this by the method in Lemma \ref{lem2} and the skills of group theory. We also def{}ine the "outward-pointing" inner product, which provides the orientation of all inner products. Finally, we examine the two matrices $S_6^{(0)}$ and $H^{(1)}$, with the results in \cite{WOS:001272956200002,bw09}, we claim that CHM containing only three distinct elements does not belong to a MUB trio.

The rest of this section is organized as follows. In Sec. \ref{sec:pre}, we introduce the def{}initions and facts used in this paper. Then we introduce our main results on CHM containing only three distinct elements and whether such CHM belongs to MUB trios in Sec. \ref{sec:res}. Finally we concluded in Sec. \ref{sec:con}.

\section{Preliminaries}
\label{sec:pre}

We follow the def{}inition in \cite{lma2020} to def{}ine the equivalence and complex equivalence.
\begin{definition}
(i) Let the monomial unitary matrix be a unitary matrix each of whose rows and columns has exactly one nonzero entry. The entry has modulus one. Let $\mathcal{M}_n$ be the set of $n\times n$ monomial unitary matrices.

(ii) Two $n\times n$ matrices U and V are complex equivalent when $U = PVQ$ where $P,Q\in\mathcal{M}_n$. If P,Q are both permutation matrices then we say that U,V are equivalent.
\end{definition}

\begin{lemma}
Suppose S is a CHM containing exactly three distinct elements, and all elements of the f{}irst row of S are one. Then S is complex equivalent to the Tao matrix.  
\end{lemma}
The above result is from \cite{WOS:001272956200002}. We want to generalize this result.
\begin{lemma}\label{lemtao}
Suppose S is a CHM containing only $\{1,\omega,\omega^2\}$, then S is complex equivalent to the Tao matrix, denoted by $S_6^{(0)}$.
\begin{eqnarray}
S_6^{(0)}=\begin{bmatrix}
1&1&1&1&1&1\\
1&1&\omega&\omega&\omega^2&\omega^2\\
1&\omega&1&\omega^2&\omega^2&\omega\\
1&\omega&\omega^2&1&\omega&\omega^2\\
1&\omega^2&\omega^2&\omega&1&\omega\\
1&\omega^2&\omega&\omega^2&\omega&1
\end{bmatrix}.
\end{eqnarray}
\end{lemma}
\begin{proof}
We can take $A\in \mathcal{M}_6$ which contains only $\{0,1,\omega,\omega^2\}$ to let $HA$ be a CHM with the property that all elements in row 1 are one. Moreover, $HA$ also contains only $\{1,\omega,\omega^2\}$, so Lemma \ref{lemtao} shows that such matrix is complex equivalent to the Tao matrix.
\end{proof}
\begin{lemma}\label{lemrank1}
If a $6 \times 6$ CHM X contains a $2\times 3$ submatrix with rank one, then X is
 complex equivalent to the matrix from the following two-parameter family
 \begin{eqnarray}
 \label{Hab}
H(\alpha,\beta)=\begin{bmatrix}
1&1&1&1&1&1\\
1&1&1&-1&-1&-1\\
1&\omega&\omega^2&\a&\a\omega&\a\omega^2\\
1&\omega&\omega^2&-\a&-\a\omega&-\a\omega^2\\
1&\omega^2&\omega&\b&\b\omega^2&\b\omega\\
1&\omega^2&\omega&-\b&-\b\omega^2&-\b\omega\\
\end{bmatrix}.
\end{eqnarray}
\end{lemma}

\begin{lemma}\label{lem1oo2}
 Suppose a $6\times6$ CHM H contains a submatrix $\begin{bmatrix}
1&1&1&1&1&1\\
1&1&\omega&\omega&\omega^2&\omega^2\\
 \end{bmatrix}$ or $\begin{bmatrix}
1&1\\
1&1\\
1&\omega\\
1&\omega\\
1&\omega^2\\
1&\omega^2
 \end{bmatrix}$ where $\omega=e^{\frac{2\pi i}{3}}$. Then H is complex equivalent to the Tao matrix or the 
 matrix from the two-parameter family $H(\a,\b)$ in Lemma \ref{lemrank1}.
\end{lemma}

Lemmas \ref{lemrank1} and \ref{lem1oo2} were obtained in \cite{WOS:001272956200002}.
\begin{lemma}\label{lemreal}
Suppose $|a|=|b|=1$. We have

(i) $a+b$ is real if and only if $a=\bar{b}$ or $a=-b$;

(ii) $ab$ is real if and only if $a=\bar{b}$ or $a=-\bar{b}$;

(iii) $a+b+ab$ is real if and only if $a=\bar{b}$ or $a=-1$ or $b=-1$;

(iv) $a+b+a\bar{b}$ is real if and only if $a=-b$ or $a=1$ or $b=-1$;

(v) $a+b+\bar{a}\bar{b}$ is real if and only if $a=\bar{b}$ or $a=1$ or $b=1$.

(vi) if also $|c|=|d|=1$ and $a+b=c+d$, then we have $a=c$ or $a=d$.
\end{lemma}
\begin{proof}
All these results can be proven easily by assuming $a=\cos\theta_1+i\sin\theta_1$,$b=\cos\theta_2+i\sin\theta_2$ and considering the coeff{}icient of $i$. Moreover, we can let $c=-a,d=-b$, so $a+b+ab=-c-d+cd$, now $c+d-cd$ is real if and only if $a=\bar{b}$ or $a=-1$ or $b=-1$, that is, $c=\bar{d}$ or $c=1$ or $d=1$. The same skill can show us the condition to let $a+b-a\bar{b},a+b+\bar{a}\bar{b}$ be real.
\end{proof}

Finally we introduce a lemma which shows some properties of MUB trios. This lemma is the result from \cite{WOS:001272956200002}.

\begin{lemma}\label{lemmub}
An MUB trio does not contain a CHM which has a $3\times3$ Hadamard submatrix.
\end{lemma}

\section{Results}
\label{sec:res}

In this section, we will take a complete classif{}ication of CHMs containing only three distinct elements. We will f{}irst classify the following three cases clearly and use the results to take a complete classif{}ication of CHMs containing only three distinct elements. Here the three cases are CHM containing three distinct elements $\{1,-1,a\},\{1,a,\bar{a}\}$ or $\{1,a,-\bar{a}\}.$ Before that we will give a useful lemma to simplify our discussion.

\begin{lemma}\label{lemh}
There does not exist a CHM $H$ of order six containing only $\{1,a,b\}$ which is complex equivalent to $H(\alpha,\beta)$ in Lemma \ref{lemrank1}.
\end{lemma}
\begin{proof}
If we do multiplications by diagonal matrices in $\mathcal{M}_6$, it is easy to f{}ind that for any two rows, the inner product of them before and after the multiplications will be the same, in the sense of multiplying by a complex number of modulus 1. 

If $b=-1$, since the inner product of the columns 4-6 of rows 1,2 in $H(\alpha,\beta)$ is -3, so that each column in columns 4-6 of the f{}irst two rows of $H$ must be the same. We shall assume the columns 4-6 of row 1 to be $[1,1,1]$. And the inner product of the columns 4-6 of rows 1,3 is $\alpha(1+\omega+\omega^2)=0$, which shows the columns 3-6 of row 3 of that CHM contains three distinct elements. But that CHM contains only $\{1,-1,a\}$, and $1-1+\bar{a}\neq 0$ shows a contradiction. We can do similar discussion for $a=-1$ or $a=-b$. In the latter case we can consider $\bar{a}H$ which contains only $\{1,-1,\bar{a}\}$.

If $a,b\neq -1$ and $a\neq -b$, the only possible $H_2$ matrices containing only $\{1,a,b\}$ are
\begin{eqnarray}\label{h2}
\begin{bmatrix}
1&a\\b&1
\end{bmatrix},\begin{bmatrix}
a&1\\b&a
\end{bmatrix},\begin{bmatrix}
b&1\\a&b
\end{bmatrix}.
\end{eqnarray}
in the sense of exchanging their rows and columns. We can obtain three relations of $a,b$ from the three matrices, they are $b=-\bar{a},b=-a^2,a=-b^2$. Since the f{}irst two rows of $H(\alpha,\beta)$ contains exactly 3 disjoint $H_2$ matrices, if they are all the same, since the inner product of the columns 4-6 of rows 1,2 
 in $H(\alpha,\beta)$ is -3, then we shall assume the columns 4-6 of rows 1,2 of $H$ to be $[1,a]$. Consider the inner product of columns 4-6 of rows 2,3, we have $1+a+b=0$. But the equation has no common solutions with one of the three relations above, so we deduce the contradiction. If the three $H_2$ submatrices contain no less than two of the three $H_2$ matrices listed above, we can solve $\{a,b\}=\{\omega,-\omega^2\}$ or $\{-\omega,\omega^2\}$. By taking conjugate of $H$, we only need to consider $\{a,b\}=\{\omega,-\omega^2\}$. We shall assume $\alpha\neq -1$, and the inner product of rows 1,3 of $H(\alpha,\beta)$ is $(1+\alpha)(1+\omega+\omega^2)=0$ (if $\alpha=-1$ we can consider the inner product of rows 1,4) and for $x,y\in\{1,\omega,-\omega^2\}$ we have $x\bar{y}\in\{1,\pm\omega,\pm\omega^2\}$, so the rows 1,3 of matrix containing only $\{1,\omega,-\omega^2\}$ must be
 \begin{eqnarray}
\begin{bmatrix}
1&1&1&1&\omega&\omega\\
1&1&\omega&\omega&1&1
\end{bmatrix}.
\end{eqnarray}
in the sense of exchanging columns. However, there are three disjoint $H_2$ matrices in rows 1,2 of the matrix, and they are all from the matrices in \eqref{h2}. Since the f{}irst row of such three $H_2$ matrices must all contain two distinct elements, but the f{}irst row of the matrix contains 1 four times, hence we deduce the contradiction. 
\end{proof}

\subsection{CHM containing only $\{1,-1,a\}$}
For $\{1,-1,a\}$, we prove a brief conclusion.
\begin{theorem}
\label{thm:(1,-1,a)}    
Suppose $H$ is a CHM of order 6 containing only $\{1,-1,a\}$. Then $H$ is complex equivalent to 
\begin{eqnarray}
\label{eq:(1,-1,a)}
H^{(1)}=\bma
i&1&1&1&1&1\\
1&i&1&1&-1&-1\\
1&1&i&-1&1&-1\\
1&1&-1&i&-1&1\\
1&-1&1&-1&i&1\\
1&-1&-1&1&1&i\\
\ema.
\end{eqnarray}
\end{theorem}
The proof will be provided in Appendix \ref{proof:(1,-1,a)}.

\begin{corollary}
\label{cr:1-1i-i}
Suppose $H$ is an order-six CHM containing only $\{1,-1,i,-i\}$. Then $H$ is complex equivalent to $H^{(1)}$ in \eqref{eq:(1,-1,a)}.  
\end{corollary}
\begin{proof}
Since we can do the right multiplication to let all the elements in row 1 in $H$ be 1, and the elements in other rows are in $\{1,-1,i,-i\}$, that is just the case we have discussed before.
\end{proof}
\begin{corollary}\label{cor1}
Suppose $H$ is a $6\times 6$ CHM containing only $\{1,a,-a\}$, then $H$ is complex equivalent to $H^{(1)}$ in \eqref{eq:(1,-1,a)}.
\end{corollary}
\begin{proof}
    Noting that $a^{-1}H$ is a CHM containing only $\{1,-1,a^{-1}\}$, by Theorem \ref{thm:(1,-1,a)} we have f{}inished our proof.
\end{proof}

Based on the previous results, we introduce a lemma which can immensely simplify our discussion in the latter subsections.
\begin{lemma}\label{lemH2}
If an $H_2$-reducible CHM of order 6 H containing only three distinct elements $\{1,a,b\}$, then it is complex equivalent to $H^{(1)}$ in \eqref{eq:(1,-1,a)}. Moreover, $a=-b$ or $-1\in\{a,b\}$.
\end{lemma}
The proof will be provided in Appendix \ref{proof:lemH2}.

\subsection{CHM containing only $\{1,a,\bar{a}\}$}
In this subsection we prove the following theorem.
\begin{theorem}\label{thm2}
Suppose $H$ is a CHM of order 6 containing only $\{1,a,\bar{a}\}$, then $H$ is complex equivalent to $S_6^{(0)}$ or $H^{(1)}$ in Lemma \ref{lemtao} and Theorem \ref{thm:(1,-1,a)}.
\end{theorem}

We will prove that by analyzing several special cases and using a valid method to deal with the general cases.

Noting that if $x,y\in\{1,a,\bar{a}\}$, then $x\bar{y}\in\{1,a,\bar{a},a^2,\bar{a}^2\}$. Then for the inner product of two rows, we denote the times $1,a,\bar{a},a^2,\bar{a}^2$ appear in the six items in this inner product by $n_1,n_2,n_3,n_4,n_5$ respectively. Then we must have
\begin{equation}\label{equa6}
n_1+n_2a+n_3\bar{a}+n_4a^2+n_5\bar{a}^2=0.
\end{equation}
We claim that $0\le n_i<3(i=1,\ldots,6)$, otherwise the matrix $H$ must have a $2\times 3$ submatrix with rank one. Then Lemmas \ref{lemrank1} and \ref{lemh} show a contradiction.
\subsubsection{Several special cases}
We f{}irst pay attention to two special cases, that is, $a=\omega$ or $a=-\omega$. For $a=\omega$, then $\{1,a,\bar{a}\}=\{1,\omega,\omega^2\}$, Lemma \ref{lemtao} shows that the matrix $H$ is complex equivalent to the Tao matrix. For $a=-\omega$, then by doing multiplications with some $A\in\mathcal{M}_6$, we f{}ind that $H$ is equivalent to a matrix with the property that all the elements of the f{}irst row are 1. This matrix contains only $\{1,\omega,\omega^2,-\omega,-\omega^2\}$. In the last paragraph of the proof of Lemma \ref{lemH2}, we know such CHM is complex equivalent to the Tao matrix. In fact, one example of a CHM containing only $\{1,-\omega,-\omega^2\}$ is the following form, denoted by $S_6^{(1)}$.
\begin{eqnarray}
\begin{bmatrix}
1&1&1&1&1&1\\
1&1&-\omega&-\omega&-\omega^2&-\omega^2\\
1&-\omega&1&-\omega^2&-\omega^2&-\omega\\
1&-\omega&-\omega^2&1&-\omega&-\omega^2\\
1&-\omega^2&-\omega^2&-\omega&1&-\omega\\
1&-\omega^2&-\omega&-\omega^2&-\omega&1\\
\end{bmatrix}.
\end{eqnarray}

So far we have done a complete analysis of these cases. Now we move on to the general cases.
\subsubsection{General cases}
We introduce a method to roughly but eff{}iciently classify all possible cases. That is, for such an array $[n_1,\ldots,n_5]$, we try to analyze just whether the left-hand side of equation (\ref{equa6}) is real. For example, for $[1,1,1,1,2]$, which means that the left-hand side of (\ref{equa6}) is $1+a+\bar{a}+a^2+2\bar{a}^2$, we can easily f{}ind that $1$, $a+\bar{a}$, and $a^2+\bar{a}^2$ are always real, that is to say, to let $1+a+\bar{a}+a^2+2\bar{a}^2$ be real, $\bar{a}^2$ must also be real. And from this we conclude that $\bar{a}^2=\pm 1$, then $a=\pm 1$ or $\pm i$, and all these cases are discussed before. This method gives us an easy way to classify all cases, let alone the solutions of original equations. We can remove some terms of the equation that are obviously real, like $a+\bar{a}$. The remaining terms of the equation are what we can not select a part of the remaining terms that must be real, and we call these remaining terms "pending terms", also, we call the amount of the remaining terms "the amount of pending terms", here the amount is calculated with multiplicity.

One important thing is that the solutions obtained from letting the pending terms be real, we call it "the solutions from the pending terms" in the following pages, are not always the solutions of the original equation, while the solutions of the original equation are always the solutions from the pending terms. That is to say, the solutions of the original equation are included in the solutions from the pending terms. Now we use this method to prove our claim below.
\begin{lemma}\label{lem2}
Suppose $H$ is a $6\times 6$ CHM containing only $\{1,a,\bar{a}\}$, then for any two distinct arrays $[n_1,\ldots,n_5]$ and $[n_1',\ldots,n_5'](0\le n_i,n_i'<3,i=1,\ldots,5)$, we call the equations with coeff{}icient from the two arrays their original equations, if their original equations are not the same and not be conjugate to each other, then the two original equations either have common but simple solutions or have no common solutions. Here simple solutions $\{1,a,\bar{a}\}$ are included in $\{1,\omega,\omega^2\},\{1,-\omega,-\omega^2\},\{1,a,-a\},\{1,-1,a\}$.
\end{lemma}

\begin{proof}
Think about a simple problem: How can we split the number 6 to no more than 5 numbers which should be non-negative and less than 3? The answer is $6=2+2+2=2+2+1+1=2+1+1+1+1$. This implies that the array $[n_1,\ldots,n_5]$ must either contain $2$ there times, or contain $2$ and $1$ both two times, or contain $2$ one time and $1$ four times. We denote these three possible cases by cases 1-3.

For case 1, the amount of pending terms can only be 0,2,4. If the amount is 0, then the original equation must be $2+2a+2\bar{a}=0$ or $2+2a^2+2\bar{a}^2=0$ or their conjugate, the solutions of these equations are $a=\pm\omega$ or $a=\pm\omega^2$, which are all included in simple solutions. If the amount is 2, then the pending terms can only be one of $2a,2\bar{a},2a^2,2\bar{a}^2$, the solutions from the pending terms are $a=-1$ or $\pm i$, which are all simple. If the amount is 4, then the pending terms must be $2(b+c)$, here $b\in\{a,\bar{a}\},c\in\{a^2,\bar{a}^2\}$, we denote the two sets by $A_1,A_2$ respectively. Lemma \ref{lemreal} shows that $2(b+c)$ is real if and only if $b=\bar{c}$ or $b=-c$. Substitute the values of $b$ and $c$ in sequence, the solutions from the pending terms are included in  $a=-1,\omega,\omega^2,-\omega$ or $-\omega^2$. All these solutions are simple.

For case 2, the amount of pending terms can only be 0,1,2,3. If the amount is 0, then the original equation must be $a+\bar{a}+2(a^2+\bar{a}^2)=0$ or $2(a+\bar{a})+a^2+\bar{a}^2=0$. The solutions of these equations are never included in simple solutions. If the amount is 1, the solutions from the pending terms are  $a=\pm 1$ or $a^2=\pm 1$, which are all simple. If the amount is 2, then the pending terms should be $b+c$ or $2d$, where $b\in A_1,c\in A_2,d\in A_1\cup A_2$. The solutions from $b+c$ or $2d$ have been discussed before and all solutions are simple. If the amount is 3, the pending terms should be $b+2c$ or $2b+c$, where $b\in A_1,c\in A_2$, In this case, we claim that all the pending terms must be $a+2a^2$ or $a+2\bar{a}^2$ or $2a+a^2$ or $2a+\bar{a}^2$ or their conjugate. By assuming $a=e^{i\theta}$, we know the solutions from $2a+a^2$ or $2a+\bar{a}^2$ are simple, and the solutions from $a+2a^2$ or $a+2\bar{a}^2$ are either $a=\pm 1$ or not simple.

For case 3, the amount of pending terms can only be 0,1. If the amount is 0, then the original equation must be $2+a+\bar{a}+a^2+\bar{a}^2=0$, the solutions of this equation are $a=\omega$ or $a=\omega^2$ or $a=\pm i$, all of these are simple. If the amount is 1, this is a discussed case and all the solutions are simple.

So we only need to consider the cases when the arrays are $[0,1,1,2,2],[0,2,2,1,1]$,\\$[1,2,1,2,0]$ or $[1,2,1,0,2]$ or their conjugate versions. We denote the four original equations from the arrays by Eq.1-4 and their conjugate versions by Eq.1$'$-4$'$. Combine two of them, we claim that there are common simple solutions or no common solutions. For Eq.1,2, consider $2\times$Eq.1$-$Eq.2; For Eq.1,3, consider Eq.3+Eq.3$'$-Eq.1; For Eq.1,4, consider Eq.4+Eq.4$'$-Eq.1; For Eq.2,3, consider $2\times$Eq.2-Eq.3-Eq.3$'$; For Eq.2,4, consider $2\times$Eq.2-Eq.4-Eq.4$'$; For Eq.3,4, consider Eq.3$-$Eq.4. Moreover, there are some cases that two different original equations have no common solutions, because the solutions from pending terms are not always the solutions from the original equation. Hence we have f{}inished the proof.
\end{proof}

Lemma \ref{lem2} inspires us that for a CHM $H$ containing only $\{1,a,\bar{a}\}$, if two inner products of different rows are neither the same nor conjugate to each other, then we claim that $H$ is complex equivalent to $S_6^{(0)}$ or $H^{(1)}$. Also, if the original equation is not Eq.1-4 or Eq.1$'$-4$'$, our claim also holds. So the remaining case is that all the original equations derived from the inner product of two distinct rows are the same or conjugate to each other and included in Eq.1-4 and Eq.1$'$-4$'$.

We use the group theory to simplify this problem. For an additive group $\langle Z_5,+\rangle$ acting on a set $A=\{1,a,\bar{a},a^2,\bar{a}^2\}$, we try to let such action representing the inner product of two elements. To realize that, we try to set up an isomorphism between $A$ and $Z_5$. We def{}ine a mapping $f$ with $f(1)=0,f(a)=1,f(\bar{a})=4,f(a^2)=2,f(\bar{a}^2)=3$, although the mapping $f$ is not a isomorphism, we have $f(a\cdot\bar{a})=f(a)+f(\bar{a})=0$,$f(a^2\cdot\bar{a}^2)=0$,$f(a\cdot a)=f(a)+f(a)=2=f(a^2)$,$f(\bar{a}^2)=2f(\bar{a})$.
We ignore the value of $f(a\cdot a^2)$ and $f(\bar{a}\cdot \bar{a}^2)$ because these values make no difference in analysing our problem. We specify that when computing the inverse mapping $f^{-1}$, we must have $f^{-1}[0,1,2,3,4]=[1,a,a^2,\bar{a}^2,\bar{a}]$.

Now we can make the group action on a set represent the inner product of two elements, moreover, for two rows $[x_1,\ldots,x_6]$,$[y_1,\ldots,y_6]$, $x_i,y_i\in A(i=1,\ldots,6)$, the inner product of the two rows can be computed in the following steps: f{}irst, compute $f[x_1,\ldots,x_6]$ and $-f[y_1,\ldots,y_6]$; Then compute $f[x_1,\ldots,x_6]-f[y_1,\ldots,y_6]$ and denote that expression by $[z_1,\ldots,z_6]$; F{}inally, we compute $\sum\limits_{i=1}^6f^{-1}(z_i)$ and this is the inner product of the two rows. We can easily verify that claim by the equation $f(x_1)-f(y_1)\equiv f(x_1)+f(\bar{y_1})\equiv f(x_1\bar{y_1})\mod 5$. In particular, we claim that $f(\bar{y})=5-f(y)$ if $y\neq 1$,$f(\bar{y})=f(y)=0$ if $y=1$.

Hence when we consider a CHM $H$ with all elements in row 1 are one and containing only elements from $A$, we can also consider $f(H)$, which contains only elements from $Z_5$ and all elements in the f{}irst row of $f(H)$ are 0.

Now consider the f{}irst three rows, by exchanging rows we shall assume them to be rows 1-3. Then we shall assume the f{}irst three rows of $f(H)$ to be:
\begin{eqnarray}
\begin{bmatrix}
0&0&0&0&0&0\\
x_1&x_2&x_3&x_4&x_5&x_6\\
y_1&y_2&y_3&y_4&y_5&y_6
\end{bmatrix}.
\end{eqnarray}

The mapping $f$ maps the left side of Eq.1-4 and Eq.1$'$-4$'$ to $15,15,8,9$ and $15,15,12,11$.  We shall assume that the inner products of row 1 and rows 2,3 are equal according to inclusion-exclusion principle, then $\sum\limits_{i=1}^6x_i=\sum\limits_{i=1}^6y_i$. Consider the inner product of rows 2,3, we have
\begin{equation}\label{equa8}
\sum\limits_{i=1}^6f(f^{-1}(x_i)\overline{f^{-1}(y_i)})\equiv 30+\sum\limits_{i=1}^6(x_i-y_i)=30
\equiv 0\mod 5.
\end{equation}
But the inner product of rows 2,3 must have the same form as other inner products, which means $\sum\limits_{i=1}^6x_i\equiv \sum\limits_{i=1}^6y_i\equiv 0\mod 5$. Hence $\sum\limits_{i=1}^6x_i$ must be 15, corresponding to $[1,2,2,3,3,4]$ or $[1,1,2,3,4,4]$ respectively. 

If the second row is $[1,2,2,3,3,4]$, then consider that the inner product of rows 2,3 has the same form and row 3 is a permutation of $[1,2,2,3,3,4]$, we can solve the third row, that is $[3,4,3,2,1,2]$. Since such claim holds for any row in rows 3-6, we can never solve the fourth row then, hence we deduce a contradiction. The contradiction can be similarly deduced for the case when the second row is $[1,1,2,3,4,4]$. Hence we have f{}inished our proof of Theorem \ref{thm2}.
\begin{corollary}\label{cor2}
Suppose $H$ is a CHM of order 6 containing only $\{1,a,a^2\}$, then $H$ is complex equivalent to $S_6^{(0)}$ or $H^{(1)}$ in Lemma \ref{lemtao} and Theorem \ref{thm:(1,-1,a)}.
\end{corollary}
\begin{proof}
Noting that $a^{-1}H$ is a CHM containing only $\{1,a,\bar{a}\}$, by Theorem \ref{thm2} we have f{}inished the proof.
\end{proof}
\subsection{CHM containing only $\{1,a,-\bar{a}\}$}
In this subsection we prove a main result about CHM containing only $\{1,a,-\bar{a}\}$.
\begin{theorem}\label{thm3}
$6\times 6$ CHM containing only $\{1,a,-\bar{a}\}$ does not exist.
\end{theorem}

We will prove that by analyzing several special cases and using a similar method in the previous subsection to deal with the general cases.

Noting that if $x,y\in\{1,a,-\bar{a}\}$, then $x\bar{y}\in\{1,a,-a,\bar{a},-\bar{a},-a^2,-\bar{a}^2\}$. Then for the
inner product of two rows, we denote the times $1,a,-a,\bar{a},-\bar{a},-a^2,-\bar{a}^2$ appear in the six items in this inner product by $n_1,n_2,n_3,n_4,n_5,n_6,n_7$ respectively. Then we must have 
\begin{equation}\label{equa9}
n_1+(n_2-n_3)a+(n_4-n_5)\bar{a}-n_6a^2-n_7\bar{a}^2=0.
\end{equation}
Lemma \ref{lemrank1},\ref{lemh} shows that $0\le n_i<3(i=1,\ldots,6)$. 
\subsubsection{Several special cases}
We f{}irst pay attention to the special cases, that is, $a=\pm\omega$, since we can take conjugate to $H$, so we only need to discuss the case $a=\omega$. The CHM $H$ containing only $\{1,\omega,-\omega^2\}$, by the proof of Lemma \ref{lemH2}, we know that such CHM never exists. So we have f{}inished our discussion of the special cases.
\subsubsection{General cases}
We introduce a lemma which is similar to Lemma \ref{lem2} and use the method by analyzing the pending terms to prove it.

\begin{lemma}\label{lem3}
Suppose $H$ is a $6\times 6$ CHM containing only $\{1,a,-\bar{a}\}$, then for any two distinct arrays $[n_1,\ldots,n_7]$ and $[n_1',\ldots,n_7'](0\le n_i,n_i'<3,i=1,\ldots,5)$, if their original equations are neither the same nor conjugate to each other, then the two original equations either have common but simple solutions or have no common solutions. Now the simple solutions are $a=\pm 1,\pm i,\pm\omega,\pm\omega^2$.
\end{lemma}
The proof of Lemma \ref{lem3} will be provided in Appendix \ref{proof:lem3}.

From the proof of Lemma \ref{lem3} we f{}ind that only when the original equation is $\pm(a+\bar{a})-2(a^2+\bar{a}^2)=0$,$\pm2(a+\bar{a})-(a^2+\bar{a}^2)=0$ or $2\pm(a+\bar{a})-(a^2+\bar{a}^2)=0$, it has no simple solutions. Then we denote $A_1=\{1,a,\bar{a},-a^2,-\bar{a}^2\}$,$A_2=\{1,-a,-\bar{a},-a^2,-\bar{a}^2\}$, and def{}ine $f_i:A_i\to Z_5$ by $f_1[1,a,\bar{a},-a^2,-\bar{a}^2]=[0,1,4,3,2]$ and $f_2[1,-a,-\bar{a},-a^2,-\bar{a}^2]=[0,1,4,3,2]$. For $n_3=n_5$ we will consider the mapping $f_1$ and for $n_2=n_4$ we will consider the mapping $f_2$, so we can do similar discussions as in the last subsection. Since the group actions are similar, so the result in the last subsection still holds, which means if the inner products of distinct rows or columns are the same or conjugate to each other and the solutions of the original equations are not simple, then such CHM does not exist. Recall that all simple solutions $a=\pm 1,a=\pm i,a=\pm\omega,a=\pm\omega^2$ correspond to CHM containing only $\{1,\omega,-\omega^2\}$ or $\{1,-\omega,\omega^2\}$, hence we have proven Theorem \ref{thm3}.
\begin{corollary}\label{cor3}
The $6\times 6$ CHM containing only $\{1,a,-a^2\}$ does not exist.
\end{corollary}
\begin{proof}
Noting that $a^{-1}H$ is a CHM containing only $\{1,-a,\bar{a}\}$, if we let $b=-a$, then by Theorem \ref{thm3} we have f{}inished the proof.
\end{proof}
\subsection{CHM containing only $\{1,a,b\}$}
Now we give a complete classif{}ication to all CHM containing only $\{1,a,b\}$. Here is our main result of this paper.

\begin{theorem}\label{thmbest}
Suppose $H$ is a $6\times6$ CHM containing only $\{1,a,b\}$, then $H$ is complex equivalent to $S_6^{(0)}$ or $H^{(1)}$ in Lemma \ref{lemtao} and Theorem \ref{thm:(1,-1,a)}. Moreover, any CHM containing only three distinct elements is not a member of MUB trio.
\end{theorem}
The proof of Theorem \ref{thmbest} will be provided in Appendix \ref{proof:thmbest}.

\section{Conclusion}
\label{sec:con}

We have taken a complete classif{}ication of CHM containing only three distinct elements. The surprising and brief result is that all such CHM can only be complex equivalent to an $H_2$-reducible matrix $H^{(1)}$ in Theorem \ref{thm:(1,-1,a)} or a non-$H_2$-reducible matrix $S_6^{(0)}$ in Lemma \ref{lemtao}. The result gives us a deeper understanding of CHMs of order 6 since the least amount of distinct elements one CHM can contain is just three. And by analyzing CHM containing a small amount of distinct elements we shall obtain some special matrices which can give us a new angle to f{}ind more non-$H_2$-reducible CHM and understand their property. Our method in proving the main result can also help us to analyze CHM of order 6 whose elements in the f{}irst row are all 1. For example, if such matrix contains exactly 7 distinct elements, and there are three pairs of pairwise conjugated elements, our method works. We will focus on such matrices containing more distinct elements and try to do a complete classif{}ication of them, in this way we may classify all the CHM of order 6 step by step.

\section*{ACKNOWLEDGMENTS}
Authors were supported by the NNSF of China (Grant No. 12471427).

\bibliographystyle{unsrt}

\bibliography{chm}

\begin{appendix}
\section{The proof of Lemma \ref{thm:(1,-1,a)}}
\label{proof:(1,-1,a)}
\begin{proof}
We f{}irst observe that for $x,y\in\{1,-1,a\}$, then  $x\bar{y}\in\{1,-1,a,-a,\bar{a},$\\
$-\bar{a}\}$. For a $2\times 6$ submatrix in $H$:
\begin{eqnarray}
\begin{bmatrix}
x_1&x_2&x_3&x_4&x_5&x_6\\
y_1&y_2&y_3&y_4&y_5&y_6
\end{bmatrix}.
\end{eqnarray}
We suppose that the element 1 appears $n_1$ times in $x_i\bar{y_i},i=1,2,\ldots,6$. Similarly, $-1,a,-a,\bar{a}$ and $-\bar{a}$ appear $n_2,n_3,n_4,n_5$ and $n_6$ times respectively. We always use such notations when computing the inner product between two rows or columns. Then we get the system of equations below.
\begin{equation}\label{equa3}
\begin{cases}
n_1-n_2+(n_3-n_4)a+(n_5-n_6)\bar{a}=0,\\
\sum\limits_{i=1}^6n_i=6,\\
n_i\geq 0,i=1,2,\ldots,6.
\end{cases}\end{equation}

To let the left-hand side of the f{}irst equation of (\ref{equa3}) be real, we must have $n_3-n_4=n_5-n_6$.We then denote $n_0=n_3-n_4$, the f{}irst equation of (\ref{equa3}) then becomes the following form:
\begin{equation}\label{equa4}
n_1-n_2+n_0(a+\bar{a})=0.
\end{equation}
Inspired by (\ref{equa4}), we categorize the discussion as follows.
\subsubsection{Classif{}ied discussion}
Case 1. $n_0=\pm 3$, Lemma \ref{lemrank1},\ref{lemh} show a contradiction.

Case 2. $n_0=\pm 2$, in consideration of symmetry, we only consider $n_0=2$, now $n_k$ can be

Subcase 21. $n_3=n_5=2,n_4=n_6=0$, the inner product is $n_1-n_2+4Re[a]=0$. It is obvious that $n_1-n_2$ can only be $0,\pm 2$. If $n_1-n_2=0$,  we have $4Re[a]=0$, which implies that $a=\pm i$. This case will be discussed later. If $n_1-n_2=\pm 2$, then $a=\pm\omega$ or $a=\pm\omega^2$. Then the f{}irst two rows will always be complex equivalent to
\begin{eqnarray}
\begin{bmatrix}
1&1&1&1&1&1\\
1&1&\omega&\omega&\omega^2&\omega^2
\end{bmatrix}.
\end{eqnarray}
Hence Lemma \ref{lem1oo2} shows that such CHM is complex equivalent to the Tao matrix or $H(\alpha,\beta)$. But both of them will never be complex equivalent to a CHM containing only $\{1,-1,a\}$.

Subcase 22. $n_3=2,n_4=0,n_5=3,n_6=1$. Now $n_1=n_2=0$, so that $a=\pm i$, which will be discussed later.

Case 3. $n_0=\pm 1$, we only need to consider $n_0=1$, then $n_k$ can be

Subcase 31. $n_3=n_5=1,n_4=n_6=0$, the inner product is $n_1-n_2+2Re[a]=0$, here $n_1-n_2$ can only be $0,\pm 2,\pm 4$. If that is 0 or $\pm 2$, we have $2Re[a]=0$ or $\pm 2$, the only cases that we need to discuss is $Re[a]=0$, which means $a=\pm i$. If that is $\pm 4$, since $Re[a]\le 1$, we deduce a contradiction.

Subcase 32. $n_3=n_5=2,n_4=n_6=1$ or $n_3=3,n_4=2,n_5=1,n_6=0$, then $n_1=n_2=0$, so $a+\bar{a}=0$, which means $a=\pm i$.

Subcase 33. $n_3=2,n_4=1,n_5=1,n_6=0$. The inner product is $n_1-n_2+a+\bar{a}=0$, here $n_1-n_2$ can only be $0$ or $\pm 2$, then $a=\pm i$ or $a=\pm1$. The only cases that we need to discuss are $a=\pm i$.

Case 4. $n_0=0$. If there exists two distinct rows or columns satisfying the condition in case 1-3, then the result in case 1-3 can solve. So we only need to consider such matrices, the inner product of any two distinct rows and columns of which satisf{}ies $n_0=0$.

We assume that the element $a$ appears in the $k^{th}$ row of the matrix $t_k$ times, $[t_1,\ldots,t_6]$ shows the times element $a$ appears in every row. Since for any two distinct rows we have $n_0=0$, so there contains three disjoint $H_2$ matrices in the two rows. If $a\neq\pm i$, such matrices can only be
\begin{eqnarray}
\begin{bmatrix}
a&a\\1&-1
\end{bmatrix},\begin{bmatrix}
a&1\\a&-1
\end{bmatrix},\begin{bmatrix}
1&1\\1&-1
\end{bmatrix}.
\end{eqnarray}
or matrices exchanging their rows and columns. So $t_k(k=1,\ldots,6)$ must all be odd or even.

If all the $t_k$ are even, for $t_k=4$ or 6, we can multiply $\bar{a}$ to the whole $k^{th}$ row to make the amount of imaginary elements in that row decrease. So we can take several multiplications to let the matrix contains no less than $4\times 6=24$ real elements. In \cite{Liang2019} Theorem 8 shows such matrix can only be complex equivalent to $H^{(1)}$, since the inner product of rows 1,3 of $M_{24}$ shows $n_0=2$.

If all the $t_k$ are odd, we claim that such $3\times 6$ submatrix which satisf{}ies $[3,3,3]$ never exists. That means the element $a$ appears in each row 3 times. Since $n_0=0$, such submatrix can only be
\begin{eqnarray}
\begin{bmatrix}
a&a&a&*&*&*\\
a&*&*&a&a&*\\
*&a&*&a&*&a
\end{bmatrix}.
\end{eqnarray}
or the submatrices exchanging rows and columns. We shall assume the submatrix to be
\begin{eqnarray}
\begin{bmatrix}
a&a&a&-K&K&K\\
a&K&-K&a&a&-k\\
*&a&*&a&*&a
\end{bmatrix}.
\end{eqnarray}
The orthogonality of rows 2,3 shows that the last row of that submatrix is $[-K,a,K,a,K,a]$. But then the inner product of rows 1,3 never equals to 0, which is a contradiction.

So there are at most two rows which contains $a$ three times. For $t_k=5$, we can multiply $\bar{a}$ to the whole $k^{th}$ row to make the amount of imaginary elements in that row decrease. So we can take several multiplications to let the matrix contains no less than $5\times 4+3\times 2=26$ real elements. In \cite{Liang2019}, Theorem 8 shows such matrix can only be complex equivalent to $H^{(1)}$.
\subsubsection{$a=\pm i$}
The only cases we have to discuss are $a=\pm i$. In this case, for a CHM $H=[h_{jk}]_{6\times 6}$, we can take $A\in\mathcal{M}_6$ to make the f{}irst row of $HA$ containing only 1. Just take
$A=\diag[h_{11}^{-1},\ldots,h_{16}^{-1}]$,
then obviously the row 1 of $HA$ is all 1. Moreover, the other elements of the matrix belongs to $\{1,-1,i,-i\}$.

F{}irst, any row in rows 2-6 never contains only $\{1,-1\}$, otherwise we can get a $2\times 3$ submatrix with rank 1. So Lemma \ref{lemrank1},\ref{lemh} gives the contradiction. Similarly, any row in rows 2-6 never contains only $\{i,-i\}$. Thus any row in rows 2-6 contains $i,-i$ totally 2 or 4 times. So we can assume every row in rows 2-6 totally contains $i,-i$ 2 times, otherwise we can multiply $i$ to the whole row containing them 4 times. In \cite{lma2020} Lemma 7(ii.d) shows the matrix $H$ has the following form:
\begin{eqnarray}
\begin{bmatrix}
1&1&1&1&1&1\\
-i&i\\
-i&&i\\
-i&&&i\\
-i&&&&i\\
-i&&&&&i
\end{bmatrix}.
\end{eqnarray}
By multiplying the column 1 by $i$, $H$ has the following form:
\begin{eqnarray}
\begin{bmatrix}
i&1&1&1&1&1\\
1&i&x_{23}&x_{24}&x_{25}&x_{26}\\
1&x_{32}&i&x_{34}&x_{35}&x_{36}\\
1&x_{42}&x_{43}&i&x_{45}&x_{46}\\
1&x_{52}&x_{53}&x_{54}&i&x_{56}\\
1&x_{62}&x_{63}&x_{64}&x_{65}&i
\end{bmatrix}.
\end{eqnarray}
Computing the inner product of rows 1-6, we shall solve all the $x_{jk}$. We can assume the second and third rows to be $[1,i,1,1,-1,-1]$ and $[1,1,i,-1,1,-1]$ since for other cases we can solve a matrix which is complex equivalent to that. So, by our computation, $H$ is complex equivalent to the following matrix, denoted by $H^{(1)}$.
\begin{eqnarray}
H^{(1)}=\begin{bmatrix}
i&1&1&1&1&1\\
1&i&1&1&-1&-1\\
1&1&i&-1&1&-1\\
1&1&-1&i&-1&1\\
1&-1&1&-1&i&1\\
1&-1&-1&1&1&i
\end{bmatrix}.
\end{eqnarray}
So we have f{}inished the proof.  
\end{proof}

\section{The proof of Lemma \ref{lemH2}}
\label{proof:lemH2}
\begin{proof}
We f{}irst consider $H_2$ matrices containing only $\{1,a,b\}$. If there exists one row or column containing only one element, we claim that $a=-b$ or $-1\in\{a,b\}$. We shall assume the $H_2$ matrix to be $\begin{bmatrix}x&x\\y&z\end{bmatrix}$. Then considering the orthogonality of the rows 1,2 we have $y=-z$, hence the claim is proven. Then Theorem \ref{thm:(1,-1,a)} and Corollary \ref{cor1} shows that a CHM of order 6 containing such submatrix is equivalent to $H^{(1)}$. So we only need to consider the cases when every row and column contain two distinct elements. In the sense of exchanging rows and columns, the only cases are 
\begin{eqnarray}
\begin{bmatrix}
1&a\\b&1
\end{bmatrix},
\begin{bmatrix}
a&1\\b&a
\end{bmatrix},
\begin{bmatrix}
b&1\\a&b
\end{bmatrix}.
\end{eqnarray}
From the three cases we can obtain the relations between $a,b$, that is $a=-\bar{b}$,$b=-a^2$,$a=-b^2$ respectively. Since $H$ contains an $H_2$ submatrix by exchanging rows and columns, and by Lemma \ref{lemreal} we know that there are three disjoint $H_2$ submatrices in rows 1,2 of $H$. If the three submatrices are the same in the sense of exchanging rows and columns, we shall assume that matrix is the f{}irst one, and we can f{}ind that each row in rows 1,2 contains 1 exactly three times. And there never exists a column of rows 1,2 containing only 1. Similarly, by exchanging rows, such result holds for rows 3,4 and rows 5,6. If there exists a $2\times 3$ submatrix containing only 1, then Lemma \ref{lemrank1},\ref{lemh} show a contradiction. Then there exists two rows as follow.
\begin{eqnarray}
\begin{bmatrix}
1&1&1&x_1&x_2&x_3\\
1&y_1&y_2&1&1&y_3
\end{bmatrix}.
\end{eqnarray}
where $x_j,y_j\in\{a,b\}$. For $\begin{bmatrix}1&x_1\\y_1&1\end{bmatrix}$, if $x_1\neq y_1$, then $\bar{y_1}+x_1=0$. If $x_1=y_1$, then the inner product of this submatrix is $x_1+\bar{x_1}$, which is real. It implies that $x_1+x_2+\bar{y_1}+\bar{y_2}$ must be real, then $x_3\bar{y_3}$ must be real, which means $x_2=\pm y_2$, which has been discussed before. If there are at least two distinct $H_2$ submatrices among the three submatrices. It is easy to f{}ind that combining any two relations among the three relations, we obtain $\{a,b\}=\{\omega,-\omega^2\}$ or $\{-\omega,\omega^2\}$. By taking conjugate of $H$, we only need to consider $\{a,b\}=\{\omega,-\omega^2\}$.

By right multiplying a matrix $A\in\mathcal{M}_6$, $HA$ can be a CHM containing only $\{1,\pm\omega,\pm\omega^2\}$ and the f{}irst row of $HA$ contains only 1. Assuming the inner product of two rows to be $n_1+(n_2-n_3)\omega+(n_4-n_5)\omega^2=0$, The equality holds if and only if $n_1=n_2-n_3=n_4-n_5$. If $n_1\neq 0$, then the inner product turns to $n_1(1+\omega+\omega^2)=0$, which means $n_1=n_2=n_4=2,n_3=n_5=0$, so such two rows is complex equivalent to $\begin{bmatrix} 1&1&1&1&1&1\\1&1&\omega&\omega&\omega^2&\omega^2\end{bmatrix}$, then Lemma \ref{lem1oo2},\ref{lemh} show a contradiction. Then $n_1=0$, since $\sum\limits_{i=1}^5n_i=6$, we have $\{n_2,n_4\}=\{1,2\}$ or $\{0,3\}$. If $\{n_2,n_4\}=\{0,3\}$ then there exists a $2\times 3$ submatrix with rank 1, so the contradiction. Consider the inner product of row 1 and rows 2-6, we shall assume the inner product of row 1 and rows 2,3 both satisfy $n_2=1,n_4=2$, which means row 2 and 3 both contain $\omega,-\omega,\omega^2,-\omega^2$ 1,1,2,2 times respectively. It implies that there are at least two columns in row 2,3 containing only $\omega^2,-\omega^2$, which means the six items of inner product of rows 2,3 contain $\pm1$. Since $-1$ never appears in such inner product, we have $n_1\neq 0$ for the inner product of rows 2,3, which is a contradiction. Hence we have f{}inished our proof.
\end{proof}

\section{The proof of Lemma \ref{lem3}}
\label{proof:lem3}
\begin{proof}
Noting that for the original equation $n_1+(n_2-n_3)a+(n_4-n_5)\bar{a}-n_6a^2-n_7\bar{a}^2=0$, the pending terms should be included in $(n_2-n_3)a+(n_4-n_5)\bar{a}-n_6a^2-n_7\bar{a}^2$. Whether that term is real is equivalent to whether $(n_2+n_8-n_3)a+(n_4+n_8-n_5)\bar{a}+n_7a^2+n_6\bar{a}^2$ is real, here $n_8=\max\{n_3,n_5\}$, we call the latter term by "modif{}ied terms". We denote the pending terms of the modif{}ied terms by $n_2'a+n_3'\bar{a}+n_6'a^2+n_7'\bar{a}^2$ and call the pending terms "modif{}ied pending terms". It is easy to f{}ind that $n_2'=0$ or $n_3'=0$ and $n_6'=0$ or $n_7'=0$. We notice that this modif{}ied pending terms are in the same form as the pending terms in the previous subsection. And also, $n_2'+n_3'+n_6'+n_7'\le 6$, so all the results in the proof of Lemma \ref{lem2} still hold here.

But the difference here is that the restrictions are loosened to $n_2'+n_3'\le 4$. So we need to consider the following cases:

If $n_2'+n_3'=4$, then $n_2=2=n_5=2,n_3=n_4=0$ or $n_3=n_4=2,n_2=n_5=0$. The solution is simple if $n_6=2$ or $n_7=2$, since we can assume $a=e^{i\theta}$ and consider $2(\pm a+\bar{a}-a^2)$ or $2(\pm a+\bar{a}-\bar{a}^2)$ or their conjugate to be real, then it is easy to f{}ind that all the solutions from these pending terms are simple by using the Lemma \ref{lemreal}. If $\{n_6,n_7\}=\{0,1\}$, we pay attention to the original equation, that must be $2a-2\bar{a}-a^2+1=0$ or $2a-2\bar{a}-\bar{a}^2+1=0$ or their conjugate. But the solutions of these equations are included in $a=\pm 1$, which are simple. If $n_6=n_7=1$, then the modif{}ied pending terms are $4a$ or $4\bar{a}$, all the solutions from them are simple.

If $n_2'+n_3'=3$, then $\{n_2,n_5\}=\{1,2\},n_3=n_4=0$ or $\{n_3,n_4\}=\{1,2\},n_2=n_5=0$. If $n_6=n_7=0$ or 1 then the modif{}ied pending terms must be $3a$ or $3\bar{a}$, and the solutions from that are simple for sure. If $n_6\neq n_7$, then the modif{}ied pending terms must be $3a+a^2$ or $3a+2a^2$ or $3\bar{a}+a^2$ or $3\bar{a}+2a^2$ or their conjugate. When the modif{}ied pending terms are $3a+a^2$ or $3\bar{a}+a^2$ or their conjugate, then the only solutions from them are $a=\pm 1$, which are included in simple solutions. When the modif{}ied pending terms are $3a+2a^2$ or $3\bar{a}+2a^2$ or their conjugate, the original equation must be $2a-\bar{a}-2a^2+1=0$ or $2a-\bar{a}-2\bar{a}^2+1=0$ or their conjugate, by solving these equations in $|a|=1$ we know that all solutions are included in $a=\pm 1$, which are simple.

If $n_2'+n_3'= 2$, the modif{}ied pending terms can only be $2a,2a+a^2,2a+\bar{a}^2,2a+2a^2,2a+2\bar{a}^2$ or their conjugate, the solutions form one of them are all simple. 

If $n_2'+n_3'=1$, the modif{}ied pending terms can only be $a,a+a^2,a+\bar{a}^2,a+2a^2,a+2\bar{a}^2$ or their conjugate. The solutions from $a,a+a^2,a+\bar{a}^2$ or their conjugate are all simple. For other cases, we have $\{n_6,n_7\}=\{0,2\},\{|n_2-n_3|,|n_4-n_5|\}=\{0,1\}$ or $\{1,2\}$. Then $n_1=1$, the original equation can only be $1+\pm a-2a^2=0$,$1\pm\bar{a}-2a^2=0$ or their conjugate. All these original equations have no solutions in $|a|=1$ since $|1\pm a|,|1\pm\bar{a}|\le 2$ and $2|a^2|=2|\bar{a}^2|=2$. 

If $n_2'+n_3'=0$, the modif{}ied terms can only be $0,a^2,2a^2$ or their conjugate. Solutions form $a^2,2a^2$ or their conjugate are all simple. If the modif{}ied term is 0, then the original equation can only be $\pm(a+\bar{a})-2(a^2+\bar{a}^2)=0$,$\pm2(a+\bar{a})-(a^2+\bar{a}^2)=0$ or $2\pm(a+\bar{a})-(a^2+\bar{a}^2)=0$. All solutions of these original equations are never simple.

we conclude that only when $n_2-n_3=n_4-n_5,n_6=n_7$ and if $n_1=0$, then $n_2-n_3=\pm 1,n_6=2$ or $n_2-n_3=\pm2,n_6=1$, if $n_1=2$, then $n_2-n_3=\pm 1,n_6=1$,  the solutions of the original equation are not simple. For two of these original equations, we can eliminate $(a^2+\bar{a}^2)$ from the equations and f{}ind that $a+\bar{a}=0$ or $\pm 2$ or $\pm\dfrac{2}{3}$ or $\pm\dfrac{4}{3}$. For $a+\bar{a}=0$ or $\pm 2$ or $\pm\dfrac{4}{3}$, they have only simple common solutions or have no solutions. For $a+\bar{a}=\dfrac{2}{3}$, the original equations are
\begin{eqnarray}
\begin{cases}
\pm 2(a+\bar{a})-(a^2+\bar{a}^2)=0,\\
2\mp (a+\bar{a})-(a^2+\bar{a}^2)=0.
\end{cases}
\end{eqnarray}
The two times of the second equation adds the f{}irst equation shows that $4-3(a^2+\bar{a}^2)-4=0$, which implies that there are no common solutions. Hence we have f{}inished our proof of Lemma \ref{lem3}.
\end{proof}

\section{The proof of Theorem \ref{thmbest}}
\label{proof:thmbest}
\begin{proof}
We will prove that in the following steps: f{}irst we will classify all the cases with or without simple solutions, then we focus on cases without simple solutions. Here we expand the meaning of simple solutions, in the following we use "simple solution" if we have $a=b$ or $a=\bar{b}$ or $a=-b$ or $a=-\bar{b}$ or $a=b^2$ or $a=-b^2$ or $b=a^2$ or $b=-a^2$ or $a=-1$ or $b=-1$. Considering Theorem \ref{thm:(1,-1,a)},\ref{thm2},\ref{thm3} and Corollary \ref{cor1},\ref{cor2},\ref{cor3}, we know that CHM containing such $\{1,a,b\}$ can only be complex equivalent to $S_6^{(0)}$ or $H^{(1)}$.

For $x,y\in\{1,a,b\}$, we have $x\bar{y}\in\{1,a,\bar{a},b,\bar{b},a\bar{b},b\bar{a}\}$, and we denote this set by $C$. We also denote $C_1=\{a,\bar{a}\},C_2=\{b,\bar{b}\},C_3=\{a\bar{b},b\bar{a}\}$ and denote the times $1,a,\bar{a}$,\\$b,\bar{b},a\bar{b},b\bar{a}$ appear in the six items in this inner product by $n_1,n_2,n_3,n_4,n_5,n_6,n_7$ respectively. Then we have
\begin{equation}\label{equa10}
n_1+n_2a+n_3\bar{a}+n_4b+n_5\bar{b}+n_6a\bar{b}+n_7b\bar{a}=0.
\end{equation}

We also claim that $0\le n_i<3(i=1,\ldots,7)$, otherwise Lemma \ref{lemrank1},\ref{lemh} show a contradiction. Now we discuss all possible values of $n_i$.
\subsubsection{Classif{}ication of cases with or without simple solutions}
If we want to split number 6 into no more than 7 numbers which should be non-negative and less than 3, the only cases are $6=2+2+2=2+2+1+1=2+1+1+1+1=1+1+1+1+1+1$, we denote these four cases by case 1-4.

For case 1, the amount of the pending term can only be 0,2,4,6. If the amount is 0, then the original equation must be $2+2a+2\bar{a}=0$ or $2+2b+2\bar{b}=0$ or $2+2a\bar{b}+2b\bar{a}=0$. The solutions of the f{}irst equation must be $a=\omega$ or $a=\omega^2$, but this implies that the second row is the permutation of $[1,1,\omega,\omega,\omega^2,\omega^2]$, then Lemma \ref{lem1oo2} shows that $H$ is equivalent to either $S_6^{(0)}$ or $H(\alpha,\beta)$, but Lemma \ref{lemh} implies that $H$ can only be equivalent to $S_6^{(0)}$, which is a simple solution. We can do the similar discussion on the latter two equations and get the same conclusion. If the amount is 2, then the pending term must be $2a$ or $2b$ or $2a\bar{b}$ or their conjugate. The solutions from the pending terms are $a=\pm1$ or $b\pm 1$ or $a=\pm b$, which are simple. If the amount is 4, then the original equation must be $2+2c+2d=0$,where $c\in C_i,d\in C_j(i\neq j)$, hence we have $\{c,d\}=\{\omega,\omega^2\}$, this case has been discussed just before. If the amount is 6, then the original equation must be $2+2c+2d=0$ or $2e+2f+2g=0$, where $c\in C_i,d\in C_j(i\neq j),e\in C_1,f\in C_2,g\in C_3$. For $2+2c+2d=0$ we have $\{c,d\}=\{\omega,\omega^2\}$, hence the solution $a,b\in\{1,\omega,\omega^2\}$, which are all simple. For $2e+2f+2g=0$, Lemma \ref{lemreal} inspires us that $a=b$ or $a=\bar{b}$ or $a=-b$ or $a=-\bar{b}$, and they are all simple solutions.

For case 2, the amount of the pending terms can only be 0,1,2,4,5. If the amount is 0, then we have $c+d+2e+2f=0$, here $c,d\in C_i;e,f\in C_j$, no simple solutions contain in this case. We denote these cases by N.1. If the amount is 1, then for any element lasting the solutions are always simple. If the amount is 2, then the pending terms must be $2c$ or $d+e$, where $c\in C,d\in C_i,e\in C_j(i\neq j)$, the former case has been discussed just before, and the latter one, recalling Lemma \ref{lemreal} we have $d=-e$ or $d=\bar{e}$, all the solutions are simple. If the amount is 4, then $n_1=2$, We f{}ind that when $\{n_5,n_6\}=\{0,2\}$ or $\{n_2,n_3\}=\{0,2\}$ or $\{n_4,n_5\}=\{0,2\}$, then the solutions are simple if and only if $n_2=n_4,n_3=n_5$ or $n_4=n_7,n_5=n_6$ or $n_2=n_6,n_3=n_7$. For other cases, the solutions are never simple, we denote these cases by N.2. If the amount is 5, then $n_1=1$, then the original equation has simple solutions if and only if $\{n_2,n_3\}=\{0,1\},n_4=n_7,n_5=n_6$ or $\{n_4,n_5\}=\{0,1\},n_2=n_6,n_3=n_7$ or $\{n_6,n_7\}=\{0,1\},n_2=n_4,n_3=n_5$. For other cases, the solutions are never simple, we denote these cases by N.3.

For case 3, the amount of the pending terms can only be 0,1,2,3. If the amount is 0, then we have $2+c+\bar{c}+d+\bar{d}=0$, where $c\in C_i,d\in C_j(i\neq j)$, then $\{c,d\}=\{\omega,\omega^2\}$, the solutions are simple. If the amount is 1, then the pending term is $c\in C$, the solutions are obviously simple. If the amount is 2, the pending term is $2c$ or $d+e$, both of them have been discussed before, all the solutions are simple. If the amount is 3, then the pending terms must be $c_1+c_2+c_3$ or $d+2e$, the former one has been discussed and the solutions are all simple, the latter one is also discussed and never has simple solutions, we denote these cases by N.4.

For case 4, the amount of the pending terms can only be 0,1. If the amount is 0, then we have $a+\bar{a}+b+\bar{b}+a\bar{b}+b\bar{a}=0$. The solutions are never simple, we denote this case by N.5. If the amount is 1, then the pending term must be $c\in C$, the solutions are all simple.

We list the above cases for which there is no simple solution(regardless of the conjugate version):

For N.1, that is \\ $[0,1,1,2,2,0,0]$,$[0,1,1,0,0,2,2]$,$[0,2,2,1,1,0,0]$,$[0,0,0,1,1,2,2]$,$[0,2,2,0,0,1,1]$,
$[0,0,0,2,2,1,1]$.

For N.2, that is\\
$[2,2,0,1,0,1,0],[2,2,0,0,1,0,1],[2,1,0,2,0,0,1],[2,0,1,2,0,1,0],[2,1,0,0,1,2,0]$,
$[2,0,1,1,0,2,0]$.

For N.3, that is\\
$[1,1,0,2,0,2,0],[1,1,0,0,2,0,2],[1,2,0,1,0,0,2],[1,0,2,1,0,2,0],[1,2,0,0,2,1,0]$,
$[1,0,2,2,0,1,0]$.

For N.4, that is\\
$[1,1,1,2,0,1,0],[1,1,1,0,2,1,0],[1,1,1,1,0,2,0],[1,1,1,0,1,2,0],[1,2,0,1,1,1,0]$,
$[1,0,2,1,1,1,0],[1,1,0,1,1,2,0],[1,0,1,1,1,2,0],[1,2,0,1,0,1,1],[1,0,2,1,0,1,1]$,
$[1,1,0,2,0,1,1],[1,0,1,2,0,1,1]$.

For N.5, that is\\
$[0,1,1,1,1,1,1]$.
\subsubsection{Total discussion about cases without simple solutions}
Before we dive into all the cases, we try to f{}ind out the connections among solutions of these 31 cases, then we can realize the extremely severe restrictions for cases N.1-N.5. We primarily pay attention to some special cases, that is, when the original equation which takes the array as the coeff{}icient has no pending terms. Such cases only appear in N.1 and N.5. We briefly denote the six cases in N.1 by N.1.1-N.1.6, the last number is determined by the order of our list. Now for N.1.1, if another case is N.1.4 or N.1.5, we can actually obtain a common solution by assuming $a=\cos\theta_1+i\sin\theta_1$ and $b=\cos\theta_2+i\sin\theta_2$. But for other cases in N.1, by using Lemma \ref{lemreal}(vi), we can only obtain simple solutions from the subtraction of the two original equations which take the arrays as the coeff{}icient,  which implies that there does not exist a common solution. For N.1.2, the cases that there exists a common solution if and only if the other case is N.1.3 or N.1.6(we do not list the cases appeared before, the same as follows). For N.1.3, such cases can only be N.1.6. For N.1.4, such cases can only be N.1.5. For N.1.5, such cases do not exist. For N.5, if another case is in N.1, then we assume $a=e^{i\theta_1},b=e^{i\theta_2}$, and consider the real part of the two equations, we have
\begin{eqnarray}
\begin{cases}
2(n_2\cos\theta_1+n_4\cos\theta_2+n_6\cos(\theta_1-\theta_2))=0,\\
\cos\theta_1+\cos\theta_2+\cos(\theta_1-\theta_2)=0.
\end{cases}
\end{eqnarray}
where $\{n_2,n_4,n_6\}=\{0,1,2\}$. If we denote $\gamma_1=\theta_1,\gamma_2=\theta_2,\gamma_3=\theta_1-\theta_2$, then the subtraction of these two equations shows that $\cos\gamma_i=\cos\gamma_j(i\neq j)$, this equation only has simple solutions, which is a contradiction. So for two distinct arrays both in N.1 or N.5, the only possible cases that they have common solutions are (N.1.1, N.1.4), (N.1.1, N.1.5), (N.1.2, N.1.3), (N.1.2, N.1.6), (N.1.3, N.1.6), (N.1.4, N.1.5).

For N.1.1, N.1.4 and N.1.5, the three equations which take these arrays as the coeff{}icient, have no common solutions, that claim is obvious once we assume $a=e^{i\theta_1},b=e^{i\theta_2}$ and combine the equation from the real part of the three equations. For N.1.2, N.1.3 and N.1.6, the three equations have no common solutions, too. We denote the six cases in the last paragraph by Ns.1-6.

For other cases, we f{}irst pay attention to the cases in N.2-N.4. The common characteristic of these cases is that the real part of their original equation has the form as $a_1+a_2\cos\theta_1+a_3\cos\theta_2+a_4\cos(\theta_1-\theta_2)=0$. And $\theta_1,\theta_2$ are the argument of $a,b$. Also, number 1 and 2 appear in $a_1,\ldots,a_4$ both two times. Then for two distinct cases in N.2-N.4, consider the system of the real parts of their original equations as the following
\begin{align}
a_1+a_2\cos\theta_1+a_3\cos\theta_2+a_4\cos(\theta_1-\theta_2)=0;\label{s1}\\
a_1'+a_2'\cos\theta_1+a_3'\cos\theta_2+a_4'\cos(\theta_1-\theta_2)=0\label{s2}.
\end{align}

It is obvious that the cardinality of the set $I=\{i|a_i=a_i',i=1,2,3,4\}$ can only be 0,2 or 4. If the cardinality is 0, then we shall assume $a_j=a_k=2$ and denote $x_1=1,x_2=\cos\theta_1,x_3=\cos\theta_2,x_4=\cos(\theta_1-\theta_2)$, then $2\times\eqref{s1}-\eqref{s2}$ equals to $3x_j+3x_k=0$, which means $x_j=-x_k$. And for any $j,k\in\{1,2,3,4\}$, the solutions of $x_j=-x_k$ are obviously simple. If the cardinality is 2, we may assume that $j,k\notin I$, then $\eqref{s1}-\eqref{s2}$ equals to $x_j-x_k=0$, and the solutions of that are simple. If the cardinality is 4, which means the real parts of the original equations of the two distinct cases are the same, then we can take conjugate and combine the
two original equations and obtain only simple solutions from the subtraction of two original equations or their conjugate. It means that such two cases have no common solutions. So for any two distinct cases in N.2-N.4, there are no common solutions.

For one case in N.2-N.4, the other case in N.5, and consider the system of the real parts of their original equations. Now $a_1'=0,a_2'=a_3'=a_4'=2$. If $a_1=1$, then $\eqref{s1}-\eqref{s2}$ shows $1-x_j=0$, where $a_j=1,j\neq 1$, which is a simple solution. If $a_1=2$, then $2\times\eqref{s1}-\eqref{s2}$ shows $4-2x_j=0$, where $a_j=2,j\neq 1$, which means the two cases have no common solutions. So one case in N.2-N.4, the other case in N.5, there are no common solutions between the two cases.

For one case in N.2-N.4, the other case in N.1, and consider the system of the real parts of their original equations. Now $a_1'=0$. Since for any $\{i,j,k\}=\{2,3,4\}$, either $x_i=x_jx_k+\sqrt{(1-x_j^2)(1-x_k^2)}$ or $x_i=x_jx_k-\sqrt{(1-x_j^2)(1-x_k^2)}$ holds. We may assume the equation \eqref{s2} to be $x_j+2x_k=0$. Then we rewrite the equation \eqref{s1} to be $a_1+(a_k-2a_j)x_k+a_i(-2x_k^2\pm\sqrt{(1-x_k^2)(1-4x_k^2)})=0$, which can be simplif{}ied to be
\begin{eqnarray}
[a_1+(a_k-2a_j)x_k-2a_ix_k^2]^2=a_i^2(1-x_k^2)(1-4x_k^2).
\end{eqnarray}

Substitute $(a_1,a_k,a_j,a_i)$ for $(1,1,2,2)$,$(1,2,1,2)$,$(1,2,2,1)$,$(2,1,1,2)$,$(2,1,2,1)$,$(2,2,$\\$1,1)$ respectively, we can get six equations about $x_k$. They are
\begin{align*}
&24x_k^3+21x_k^2-6x_k-3=0,\\
&-12x_k^2-3=0,\\
&8x_k^3+5x_k^2-4x_k=0,\\
&8x_k^3+5x_k^2-4x_k=0,\\
&12x_k^3+6x_k^2-12x_k+3=0,\\
&-3x_k^2+3=0.
\end{align*}

The f{}irst equation has solutions $x_k=-1,\dfrac{1\pm\sqrt{33}}{16}$, here $x_k=-1$ is a simple solution. For $x_k=\dfrac{1\pm\sqrt{33}}{16}$, consider $x_j=-2x_k$ and equation (\ref{s1}) we know $x_i=\dfrac{-13\mp 3\sqrt{33}}{32}$, but neither  $x_i=x_jx_k+\sqrt{(1-x_j^2)(1-x_k^2)}$ nor $x_i=x_jx_k-\sqrt{(1-x_j^2)(1-x_k^2)}$  holds, which shows that there are no common solutions. The second equation has no solutions. The third and fourth equations both have solutions $x_k=0,\dfrac{-10\pm3\sqrt{17}}{16}$, here $x_k=0$ is a simple solution. Consider $x_j=-2x_k$ and equation (\ref{s1}) and relations of $x_i$ and $x_j,x_k$, we claim that there are no common solutions. The f{}ifth equation has solutions $x_k=\dfrac{1}{2},\dfrac{-1\pm\sqrt{3}}{2}$. For $x_k=\dfrac{1}{2}$ we have $x_j=-1$, which is simple. For $x_k=\dfrac{-1\pm\sqrt{3}}{2}$,then $x_j=1\mp\sqrt{3}$. Since $|x_j|\le 1$, then $x_k=\dfrac{-1+\sqrt{3}}{2}$ and $x_j=1-\sqrt{3}$. Then $x_i=\dfrac{7-3\sqrt{3}}{2}$, but then the relations of $x_i$ and $x_j,x_k$ never hold, so there are no common solutions. The sixth equation has solutions $x_k=\pm 1$, which are all simple.

So far, our previous discussion shows that for two distinct cases in N.1-N.5, one of them in N.2-N.4, there are no common solutions. So only when the two cases are Ns.1-6, they have common solutions.

We construct a group acting on the set $\{1\}\cup C$ then. f{}irst we choose a group $\langle Z_7,+\rangle$, then def{}ine a mapping $f$ satisfying $f(1)=0,f(a)=1,f(\bar{a})=6,
f(b)=5,f(\bar{b})=2,f(a\bar{b})=3,f(b\bar{a})=4$. Although such a mapping is not isomorphic, but the relation $f(cd)=f(c)+f(d)\mod 7,(c,d\in \{1\}\cup C$ holds if such multiplication is permitted.

Then we can do right multiplications to the CHM $H$ to let all the elements of row 1 be one. And we consider $f(H)$, then.

We consider a class of inner product, called inner product with f{}ixed orientation. For $H=[X_1,\ldots,X_6]^T$, we call an inner product $X_i\bar{X_j}$ is "outward-pointing" if $1\le i<j\le 6$. We pay attention to all outward-pointing inner products. Since any outward-pointing inner product has an original equation between $n_1+n_2a+n_3\bar{a}+n_4b+n_5\bar{b}+n_6a\bar{b}+n_7b\bar{a}=0$ and $n_1+n_3a+n_2\bar{a}+n_5b+n_4\bar{b}+n_7a\bar{b}+n_6b\bar{a}=0$, here the latter is conjugate to the former. We claim that there are three rows, the three outward-pointing inner products have the common original equation. This claim can be easily proven by $R(3,3)=6$, here $R(3,3)$ represents a Ramsey number, which means the least number of vertices in a graph that guarantees a monochromatic complete subgraph.

If we think of $X_1,\ldots,X_6$ as six different vertices that generate a complete graph, then every side that connects two vertices represents one outward-pointing inner product, which can be one of the two forms. We regard the two different forms as two colours. So our claim is changed to be: whether 6 vertices in a graph guarantees a monochromatic complete subgraph. That claim is true because $R(3,3)=6$, hence we f{}inish the proof of our claim.

For Ns.1-6, we observe that the conjugate of the original equation in N.1.i, here $i\in\{1,\ldots,6\}$, is actually the same as the original equation. So for Ns.j ($j\in\{1,\ldots,6\}$), that is (N.1.k, N.1.s), we can regard the equation in N.1.k as red colour, and the equation in N.1.s as blue colour. We claim that there must be two rows of $f(H)$, denoted by $[x_1,\ldots,x_6]$ and $[y_1,\ldots,y_6]$, the three kind of outward-pointing inner product of row 1 and the two rows are equal, that is the outcome of $R(3,3)=6$. We shall assume such three rows are rows 1-3. Then we have $\sum\limits_{i=1}^6(x_i-y_i)\equiv \sum\limits_{i=1}^6(-x_i)\equiv\sum\limits_{i=1}^6(-y_i)\mod 7$, which means $\sum\limits_{i=1}^6(x_i-y_i) \equiv 0\mod 7$,
then we f{}ind that $\sum\limits_{i=1}^6x_i\equiv 6\times7-\sum\limits_{i=1}^6(-x_i)\equiv 0\mod 7$.

It is easy to f{}ind that all the arrays in N.1 satisfy the condition that the addition of the according arrays in $f(H)$ can be divided by 7. If the three outward-pointing inner product is the original equation in N.1.1, then consider the submatrix of $f(H)$
\begin{eqnarray}
\begin{bmatrix}
0&0&0&0&0&0\\
1&2&2&5&5&6
\end{bmatrix}.
\end{eqnarray}
the row 3 is also the permutation of $[1,2,2,5,5,6]$, however, if an element in row 2 adds $-5$, the value of $f(\bar{H})$ in row 3, is in the set $\{1,2,5,6\}$, then the element must be 6. But $-5$ appears in $f(\bar{H})$ in row 3 two times, and row 2 only contains 6 one times, hence the outward-pointing inner product of rows 2,3 never equals to that of rows 1,2. Hence the contradiction.

For N.1.2, row 2 or 3 is the permutation of $[1,3,3,4,4,6]$. An element in row 2 adds $-4$, the value of $f(\bar{H})$ in row 3, belongs to $\{1,3,4,6\}$ if and only if the element is 1 or 3. The addition is $4$ or $6\mod 7$. Since 6 appears only one time in row 2, so that $f(H)$ must have the following part
\begin{eqnarray}
\begin{bmatrix}
0&0&0&0&0&0\\
1&3&3&4&4&6\\
4&4
\end{bmatrix}.
\end{eqnarray}
Then the element 6 in row 3 must be in column 3, but then the element 3 in row 3 has no suitable location. Hence the contradiction.

For N.1.3, consider the location of two element 1 in row 3, we know that $f(H)$ have the following part
\begin{eqnarray}
\begin{bmatrix}
0&0&0&0&0&0\\
1&1&2&5&6&6\\
&&1&1
\end{bmatrix}.
\end{eqnarray}
Then the element 5 in row 3 must be in column 4, but then 6 has no suitable location.

For N.1.4, check the element 3 in row 3, the only suitable location of that is the column that row 2 contains 5. The amount of element 3 and 5 lead to a contradiction.

For N.1.5, check the element 1 in row 3, the only suitable location of that is the column that row 2 contains 4. The amount of element 1 and 4 lead to a contradiction.

For N.1.6, consider the location of two element 2 in row 3, we know that $f(H)$ have the following part
\begin{eqnarray}
\begin{bmatrix}
0&0&0&0&0&0\\
2&2&3&4&5&5\\
&&&2&&2
\end{bmatrix}.
\end{eqnarray}
Then the element 3 in row 3 must be in column 3, but the 2 has no suitable location.

For N.2-N.5, we claim that there must be two rows of $f(H)$, denoted by $[x_1,\ldots,x_6]$ and $[y_1,\ldots,y_6]$, the three kind of outward-pointing inner product of row 1 and the two rows are equal, that is the outcome of $R(3,3)=6$. Hence we also have 
$\sum\limits_{i=1}^6x_i\equiv0\mod 7$, the cases in N.2-5 that satisfy the condition that the addition of the according arrays in $f(H)$ can be divided by 7 are 

$[1,1,1,0,2,1,0],[1,1,0,1,1,2,0],[1,2,0,1,0,1,1],[0,1,1,1,1,1,1]$.

By taking conjugate of $H$, here the elements in row 1 in $H$ have been modif{}ied to 1, then by inclusion-exclusion principle $f(H)$ must contain three rows in rows 2-6 which are all permutations of the same array among $[0,1,2,2,3,6],[0,1,2,3,3,5],[0,1,1,3,4,5],
[1,2,3,4,5,6]$. We shall assume that such three rows are the rows 2-4. 

If the array is $[0,1,2,2,3,6]$, then we assume the f{}irst two rows of $f(H)$ are
\begin{eqnarray}
\begin{bmatrix}
0&0&0&0&0&0\\
0&1&2&2&3&6\\
\end{bmatrix}.
\end{eqnarray}
By our computation (pay attention to the location of 0 and 1 and how can we obtain 2 or 5 in the inner product) we know that the third row of $f(H)$ can only be $[2,0,2,3,6,1]$ or $[6,2,2,0,1,3]$ or $[1,6,2,0,2,3]$ or the version exchanging the column 3 and 4, but however we choose two of them, the inner product of them is never the permutation of $[0,1,2,2,3,6]$ or $[0,1,4,5,5,6]$, we can deduce the contradiction.

If the array is $[0,1,2,3,3,5]$, then we assume the f{}irst two rows of $f(H)$ are
\begin{eqnarray}
\begin{bmatrix}
0&0&0&0&0&0\\
0&1&2&3&3&5\\
\end{bmatrix}.
\end{eqnarray}
By our computation (pay attention to the location of 0 and 1 and how can we obtain 4 or 3 in the inner product) we know that the third row of $f(H)$ can only be $[3,2,5,3,1,0]$ or $[3,2,0,5,3,1]$ or $[2,5,1,0,3,3]$ or $[5,3,1,0,3,2]$ or the version exchanging the column 4 and 5. However, for any two of them, the inner product of them is never the permutation of $[0,1,2,3,3,5]$ or $[0,2,4,4,5,6]$, which is a contradiction.

If the array is $[0,1,1,3,4,5]$, then we assume the f{}irst two rows of $f(H)$ are
\begin{eqnarray}
\begin{bmatrix}
0&0&0&0&0&0\\
0&1&1&3&4&5\\
\end{bmatrix}.
\end{eqnarray}
By our computation (pay attention to the location of 0 and 1 and how can we obtain 4 in the inner product) we know that the third row of $f(H)$ can only be $[1,4,1,0,5,3]$ or $[1,5,1,4,0,3]$ or $[3,0,1,5,1,4]$ or $[4,0,1,5,3,1]$ or the version exchanging the column 2 and 3. However, for any two of them, the inner product of them is never the permutation of $[0,1,1,3,4,5]$ or $[0,2,3,4,6,6]$, which is a contradiction.

If the array is $[1,2,3,4,5,6]$, then the inner product of every two distinct rows must be $a+\bar{a}+b+\bar{b}+a\bar{b}+b\bar{a}$. Notice that for $x,y\in\{1,a,b\}$, we have  $x\bar{y}=a,\bar{a},b,\bar{b},a\bar{b}$ or $b\bar{a}$ if and only if $(x_i,y_i)$ is equal to $(a,1),(1,a),(b,1),(1,b),(a,b)$ or $(b,a)$. So we shall assume the rows 1,2 of this matrix to be
\begin{eqnarray}
\begin{bmatrix}
1&1&a&a&b&b\\
a&b&1&b&1&a
\end{bmatrix}.
\end{eqnarray}
For any row in rows 3-6, the columns 1,2 of such rows must be $[a,b]$ or $[b,a]$, otherwise $n_3=n_5=1$ will never satisfy when considering the inner product of row 1 and that row. So by inclusion-exclusion principle we know that one of the $[a,b]$ and $[b,a]$ must appear at least three times in the columns 1,2 of rows 2-6. But then we have a $3\times 2$ submatrix which has two totally the same columns. Lemma \ref{lemrank1},\ref{lemh} shows the contradiction.

So far we have discussed all the cases with no simple solutions. So combining our result in all the previous subsections in the conclusion section, we can f{}inally claim that, CHM containing only $\{1,a,b\}$ must be complex equivalent to $S_6^{(0)}$ or $H^{(1)}$. With the Lemma \ref{lemmub}, we know that $S_6^{(0)}$ can never belong to a MUB trio. And in \cite{bw09}, in Sec.IV subsection A, we know that $H^{(1)}$ also can never belong to a MUB trio. Hence we claim that every CHM containing only three distinct elements does not belong to a MUB trio. So we have f{}inished our proof of Theorem \ref{thmbest}.
\end{proof}

\end{appendix}

\end{document}